\documentclass[aps,twocolumn]{revtex4}
\usepackage{amsmath,amssymb}
\usepackage[pdftex]{graphicx} % Allows graphics in pdf format. Removed amssymb, which clashed with teubner in defining digamma
\usepackage{epstopdf} % Allows conversion of .eps files to .pdf files
\DeclareGraphicsExtensions{.pdf,.eps,.png,.jpg,.jpeg,.mps} % Declares the various extension format allowed
\usepackage{epsfig} 
\usepackage{epic}
\usepackage{hyperref} 
\usepackage{color}
\usepackage{fancyhdr}
\usepackage{bm}
\usepackage[lofdepth,lotdepth]{subfig}
\usepackage{lipsum}

\graphicspath{{gfx/}}

\def\at{a^\dagger}
\def\bt{b^\dagger}

\begin{document} 

\title{Probing matter-field and atom-number correlations in optical lattices by global nondestructive addressing} 
\author{W. Kozlowski}
\author{S. F. Caballero-Benitez} 
\author{I. B. Mekhov}
\affiliation{Department of Physics, Clarendon Laboratory, University of Oxford, Parks Road, Oxford OX1 3PU, United Kingdom}

\begin{abstract}
	We show that light scattering from an ultracold gas reveals not only density correlations, but also matter-field interference at its
	shortest possible distance in an optical lattice, which defines key properties such as tunnelling and matter-field phase gradients. This
	signal can be enhanced by concentrating probe light between lattice sites rather than at density maxima. As addressing between two
	single sites is challenging, we focus on global nondestructive scattering, allowing probing order parameters, matter-field quadratures
	and their squeezing. The scattering angular distribution displays peaks even if classical diffraction is forbidden and we derive
	generalized Bragg conditions. Light scattering distinguishes all phases in the Mott insulator - superfluid - Bose glass phase
	transition.
\end{abstract}

\maketitle

%%%%%%%%%%%%%%%%%%%%%%%%%%%%%%%%%%%%%%%%%%%%%%%%%%%%%%%%%%%%%%%%%%%%%%%%%%%%%
%%%%%%%%%%%%%%%%%%%%%%%%%%%%%%%%%%%%%%%%%%%%%%%%%%%%%%%%%%%%%%%%%%%%%%%%%%%%%
%%
%% \section{Introduction}
%%
%%%%%%%%%%%%%%%%%%%%%%%%%%%%%%%%%%%%%%%%%%%%%%%%%%%%%%%%%%%%%%%%%%%%%%%%%%%%%
%%%%%%%%%%%%%%%%%%%%%%%%%%%%%%%%%%%%%%%%%%%%%%%%%%%%%%%%%%%%%%%%%%%%%%%%%%%%%

\section{Introduction}

The modern field of ultracold gases is successful due to its interdisciplinarity \cite{Bloch2007, Lewenstein2007}. Originally condensed
matter effects are now mimicked in controlled atomic systems finding applications in areas such as quantum information processing
(QIP). A really new challenge is to identify novel phenomena which were unreasonable to consider in condensed matter, but will become
feasible in new systems. One such direction is merging quantum optics and many-body physics \cite{Mekhov2012,RitschRMP}. The former
describes delicate effects such as quantum measurement and state engineering, but for systems without strong many-body correlations
(e.g. atomic ensembles). In the latter, decoherence destroys these effects in conventional condensed matter. Due to recent experimental
progress, e.g. Bose-Einstein condensates (BEC) in cavities \cite{EsslingerNat2010,HemmerichScience2012, ZimmermannPRL2014}, quantum optics of quantum gases can close this gap.

Here we develop a method to measure properties of ultracold gases in optical lattices (OL) by light scattering. Recent quantum
non-demolition (QND) schemes \cite{Mekhov2007, PolzikNatPh2008, DeChiaraPRA2014} probe density fluctuations, thus inevitably destroy information about phase, i.e. the conjugate variable, and as a consequence destroy matter-field coherence. In contrast, we focus on probing the atom interference between lattice sites. Our scheme is nondestructive in contrast to totally destructive methods such as time-of-flight. It enables in-situ probing of the matter-field coherence at its shortest possible distance
in an OL, i.e. the lattice period, which defines key processes such as tunnelling, currents, phase gradients, etc. This is achieved by
concentrating light between the sites. By contrast, standard destructive time-of-flight measurements deal with far-field interference
and a relatively near-field one was used in Ref. \cite{KetterlePRL2011}. Such a counter-intuitive configuration may affect works on
quantum gases trapped in quantum potentials \cite{Mekhov2008,Mekhov2012,LarsonPRL2008,MeystrePRA2009,MorigiPRL2013, Ivanov2014, Santiago2015} and quantum measurement-induced preparation of many-body atomic states \cite{Mekhov2009, Pedersen2014, Elliott2015, Mazzucchi2015}. Within the mean-field (MF) treatment, this enables probing the order parameter, matter-field quadratures and squeezing. This can have an impact on atom-wave metrology and information processing in areas where quantum optics already made progress, e.g., quantum imaging with pixellized sources of non-classical light \cite{GolubevPRA2010,KolobovRMP1999}, as an OL is a natural source of multimode nonclassical matter waves. The scattering angular distribution is nontrivial, even when classical diffraction is forbidden. We derive generalized Bragg conditions and give parameters for the only two relevant experiments to date \cite{Weitenberg2011, KetterlePRL2011}. The method works beyond MF, which we support by distinguishing all phases in the Mott insulator (MI) - superfluid (SF) - Bose glass (BG) phase transition in a 1D disordered OL. We underline that transitions in 1D are much more visible when changing an atomic density rather than for fixed-density scattering. It was only recently that an experiment distinguished a MI from a BG \cite{Derrico2014}.

%%%%%%%%%%%%%%%%%%%%%%%%%%%%%%%%%%%%%%%%%%%%%%%%%%%%%%%%%%%%%%%%%%%%%%%%%%%%%
%%%%%%%%%%%%%%%%%%%%%%%%%%%%%%%%%%%%%%%%%%%%%%%%%%%%%%%%%%%%%%%%%%%%%%%%%%%%%
%%
%%\section{The Model}
%%
%%%%%%%%%%%%%%%%%%%%%%%%%%%%%%%%%%%%%%%%%%%%%%%%%%%%%%%%%%%%%%%%%%%%%%%%%%%%%
%%%%%%%%%%%%%%%%%%%%%%%%%%%%%%%%%%%%%%%%%%%%%%%%%%%%%%%%%%%%%%%%%%%%%%%%%%%%%

\section{The Model}

\begin{figure}[htbp!]
	\centering
		\includegraphics[width=0.95\linewidth]{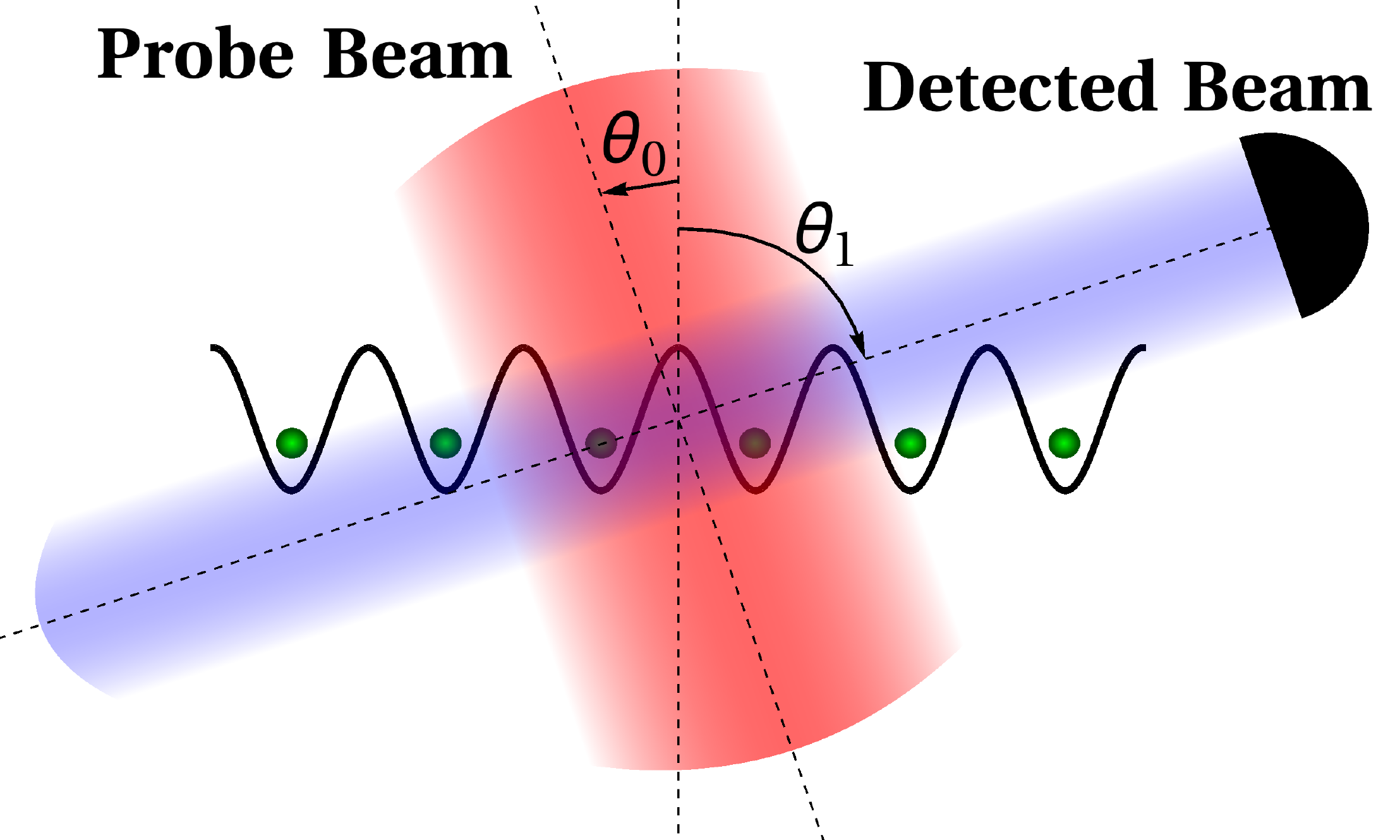}
	\captionsetup{justification=centerlast,font=small}
	\caption{(Color online) Setup. Atoms in an optical lattice are illuminated by a probe beam. The light scatters in free space or into a cavity and is measured by a detector.}
	\label{fig:Setup}
\end{figure}

The theory is based on the Bose-Hubbard (BH) model generalized for quantum light \cite{Mekhov2007a}. We consider off-resonant probe and detected light at angles $\theta_{0,1}$ with $N$ atoms at $M$ sites, $K$ of which are illuminated (Fig. \ref{fig:Setup}). Light modes
can be in free space or cavities. The Hamiltonian is
	\begin{multline}
	\label{eq:H}
		\hat{H} = \hat{H}_\text{BH} + \sum_l \hbar \omega_l \at_l a_l + \hbar\sum_{l,m} U_{lm} \at_l a_m \hat{F}_{lm}, \\
		\hat{H}_\text{BH} = - J^\text{cl} \sum^M_{\langle i, j \rangle}  \bt_i b_j + \frac{U}{2} \sum^M_i \hat{n}_i (\hat{n}_i - 1)  - \mu 				\sum^M_i \hat{n}_i,  \nonumber
	\end{multline}
$\langle i, j \rangle$ gives summation over nearest sites, $a_l$ ($l=0$,$1$) are the annihilation operators of the light modes with
frequency $\omega_l$, atom-light coupling constant $g_l$ ($U_{lm}=g_lg_m/\Delta_a$), light-atom detuning \mbox{$\Delta_a = \omega_1 - \omega_a$}. $b_i$ ($\hat{n}_i$) is the atom annihilation (number) operator with hopping amplitude $J^\text{cl}$, interaction strength
$U$, and chemical potential $\mu$. The atomic operator is $\hat{F}_{lm} = \hat{D}_{lm} + \hat{B}_{lm}$,
	\begin{equation}
	\label{eqDB}
  		\hat{D}_{lm} = \sum_{i=1}^K J^{lm}_{i,i} \hat{n}_i,
		\,\,\,\,
  		\hat{B}_{lm} = \sum_{\langle i,j \rangle}^K J^{lm}_{i,j} \bt_i b_j,
	\end{equation}
comes from overlaps of light mode functions $u_l({\bf r})$ and density operator $\hat{n}({\bf r})=\hat{\Psi}^\dag({\bf r})\hat{\Psi}({\bf r})$, after the matter-field operator is expressed via Wannier functions: $\hat{\Psi}({\bf r})=\sum_i b_i w({\bf r}-{\bf r}_i)$. $\hat{D}_{lm}$ sums the density contributions $\hat{n}_i$, while $\hat{B}_{lm}$ sums the matter-field interference terms. $J^{lm}_{i,j}$ are the convolutions of Wannier and light mode functions and are given by
	\begin{equation}
	\label{Jcoeff}
		J^{lm}_{i,j} = \int w({\bf r} - {\bf r}_i) u^*_l({\bf r}) u_m({\bf r}) w({\bf r} - {\bf r}_j)\mathrm{d}{\bf r}. 
	\end{equation}
This equation encapsulates the simplicity and flexibility of the measurement scheme that we are proposing. The operators given by Eq. (\ref{eqDB}) are entirely determined by the values of these coefficients and despite its simplicity, this is sufficient to give rise to a host of interesting phenomena via measurement back-action such as the generation of multipartite entangled spatial modes in an optical lattice \cite{Elliott2015,Mekhov2009a},  the appearance of long-range correlated tunnelling capable of entangling distant lattice sites, and in the case of fermions, the break-up and protection of strongly interacting pairs \cite{Mazzucchi2015}. Additionally, these coefficients are easy to manipulate experimentally by adjusting the optical geometry via the light mode functions $u_l({\bf r})$.

Here, we will use the fact that the light is sensitive to the atomic quantum state due to the coupling of the optical and matter fields via operators according to Eq. (\ref{eqDB}) in order to develop a method to probe the properties of an ultracold gas. Therefore, we neglect the measurement back-action and we will only consider expectation values of light observables. Since the scheme is nondestructive (in some cases, it even satisfies the stricter requirements for a QND measurement \cite{Mekhov2007a, Mekhov2012}) and the measurement only weakly perturbs the system, many consecutive measurements can be carried out with the same atoms without preparing a new sample. Again, we will show how the extreme flexibility of the the measurement operator $\hat{F}$ allows us to probe a variety of different atomic properties in-situ ranging from density correlations to matter-field interference.

For a coherent beam probe which has a negligible effect on atomic dynamics, the stationary light amplitude $a_1$ is given by $\hat{F}_{10}$ \cite{Mekhov2007a} (we drop the subscripts in $\hat{F}$, $\hat{D}$, and $\hat{B}$). In a cavity with the decay rate $\kappa$ and probe-cavity detuning $\Delta_p$, 
	\begin{equation}
		a_1 = \frac{U_{10} a_0} {\Delta_p + i \kappa} \hat{F} = C \hat{F}.
	\end{equation} 
In free space, the electric field operator in the far-field point $r$ is 
	\begin{equation}
		\hat{E}_1= \frac{\omega^2_ad_A^2E_0} {8\pi\hbar\epsilon_0c^2\Delta_a r} \hat{F} = C_E\hat{F},
	\end{equation}
where $d_A$ is the dipole moment and $E_0$ is probe electric field \cite{Scully}.

The light quadrature operators $\hat{X}_\phi = (a_1 e^{-i \phi} + \at_1 e^{i \phi})/2$  can be expressed via $\hat{F}$ quadratures, $\hat{X}^F_\beta$, 
	\begin{equation}
		\hat{X}_\phi = |C| \hat{X}^F_\beta = |C|(\hat{F}e^{-i\beta} + \hat{F}^\dagger e^{i \beta})/2, 
	\end{equation} 
where $\beta = \phi - \phi_C$, $C = |C|\exp(i\phi_C)$, and $\phi$ is the local oscillator phase . The means of amplitude and
quadrature, $\langle a_1\rangle$ and $\langle\hat{X}_\phi\rangle$, only depend on atomic mean values. In contrast, the means of light
intensity $\langle a_1^\dag a_1\rangle=|C|^2\langle\hat{F}^\dag\hat{F}\rangle$ and quadrature variance 
	\begin{equation}
		(\Delta X_ \phi)^2 = \langle \hat{X}_\phi^2 \rangle - \langle \hat{X}_\phi \rangle^2 = 1/4 + |C|^2 (\Delta X^F_\beta)^2
	\end{equation}
reflect atomic correlations and fluctuations, which is our main focus. Alternatively, one can measure the light intensity, where the ``quantum addition'' to light due atom quantum fluctuations (classical diffraction signal is subtracted), $R=\langle a^\dagger a \rangle - |\langle a \rangle|^2$, behaves similarly to $(\Delta X^F_\beta)^2$.

%%%%%%%%%%%%%%%%%%%%%%%%%%%%%%%%%%%%%%%%%%%%%%%%%%%%%%%%%%%%%%%%%%%%%%%%%%%%%
%%%%%%%%%%%%%%%%%%%%%%%%%%%%%%%%%%%%%%%%%%%%%%%%%%%%%%%%%%%%%%%%%%%%%%%%%%%%%
%%
%%\section{QND measurement of the density operator}
%%
%%%%%%%%%%%%%%%%%%%%%%%%%%%%%%%%%%%%%%%%%%%%%%%%%%%%%%%%%%%%%%%%%%%%%%%%%%%%%
%%%%%%%%%%%%%%%%%%%%%%%%%%%%%%%%%%%%%%%%%%%%%%%%%%%%%%%%%%%%%%%%%%%%%%%%%%%%%

\section{Global Nondestructive Measurement}

\subsection{On-site density measurements}

Typically, the dominant term in $\hat{F}$ is the density-term $\hat{D}$, rather than inter-site matter-field interference $\hat{B}$
\cite{Mekhov2007a,MorigiPRA2010Scat,TrippenbachPRA2009, RuostekoskiPRL2009, MekhovLP09}, because the Wannier functions' overlap is small. Our aim is to enhance the $\hat{B}$-term in light scattering by suppressing the density signal. To clarify typical light scattering, we start
with a simpler case when scattering is faster than tunneling and $\hat{F}=\hat{D}$. This corresponds to a QND scheme \cite{Mekhov2007,
Mekhov2007a, PolzikNatPh2008,DeChiaraPRA2014}. The density-related measurement destroys some matter-phase coherence in the conjugate variable \cite{Mekhov2009a,MekhovLP10,MekhovLP11} $b^\dag_ib_{i+1}$, but this term is neglected.

\begin{figure}[htbp!]
	\begin{center}
		\includegraphics[width=\linewidth]{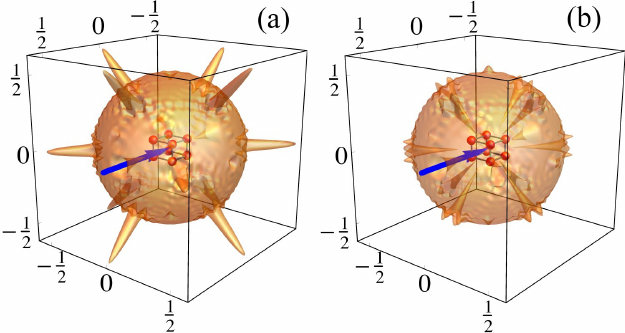}
	\end{center}
	\captionsetup{justification=centerlast,font=small}
	\caption{(Color online) Light intensity scattered into a standing wave mode from a SF in a 3D lattice (units of $R/N_K$). Arrows denote incoming travelling wave probes. The Bragg condition, $\Delta {\bf k} = {\bf G}$, is not fulfilled, so there is no classical diffraction, but intensity still shows multiple peaks, whose heights are tunable by simple phase shifts of the optical beams: (a) $\varphi_1=0$; (b) $\varphi_1=\pi/2$. Interestingly, there is also a significant uniform background level of scattering which does not occur in its classical counterpart. }
	\label{fig:Scattering}
\end{figure}

In a deep lattice, 
	\begin{equation}
		\hat{D}=\sum_i^K u_1^*({\bf r}_i) u_0({\bf r}_i) \hat{n}_i,
	\end{equation}
which for travelling [$u_l({\bf r})=\exp(i{\bf k}_l{\bf r}+i\varphi_l)$] or standing [$u_l({\bf r})=\cos({\bf k}_l{\bf r}+\varphi_l)$] waves is just a density Fourier transform at one or several wave vectors $\pm({\bf k}_1\pm {\bf k}_0)$. The quadrature for two travelling waves is reduced to 
	\begin{equation}
		\hat{X}^F_\beta=\sum_i^K \hat{n}_i\cos[({\bf k}_1-{\bf k}_2) \cdot {\bf r_i}-\beta].
	\end{equation}
Note that different light quadratures are differently coupled to the atom distribution, hence varying local oscillator phase and detection angle, one scans the coupling from maximal to zero. An identical expression exists for $\hat{D}$ for a standing wave, where $\beta$ is replaced by $\varphi_l$, and scanning is achieved by varying the position of the wave with respect to atoms. Thus, variance $(\Delta X^F_\beta)^2$ and quantum addition $R$, have a non-trivial angular dependence, showing more peaks than classical diffraction and the peaks can be tuned by the light-atom coupling.

Fig. \ref{fig:Scattering} shows the angular dependence of $R$ for standing and travelling waves in a 3D OL. The isotropic background
gives the density fluctuations [$R=K( \langle\hat{n}^2\rangle-\langle\hat{n}\rangle^2)/2$ in MF with inter-site correlations
neglected]. The radius of the sphere changes from zero, when it is a MI with suppressed fluctuations, to half the atom number at $K$ sites, $N_K/2$, in the deep SF. There exist peaks at angles different than the classical Bragg ones and thus, can be observed without being masked by classical diffraction. Interestingly, even if 3D diffraction \cite{KetterlePRL2011} is forbidden (Fig. \ref{fig:Scattering}), the peaks
are still present. As $(\Delta X^F_\beta)^2$ and $R$ are squared variables, the generalized Bragg conditions for the peaks are $2
\Delta {\bf k} = {\bf G}$ for quadratures of travelling waves, where $\Delta {\bf k}={\bf k}_0 - {\bf k}_1$ and ${\bf G}$ is the reciprocal lattice vector, and $2 {\bf k}_1 = {\bf G}$ for standing wave $a_1$ and travelling $a_0$, which is clearly different from the classical
Bragg condition $\Delta {\bf k} = {\bf G}$. The peak height is tunable by the local oscillator phase or standing wave shift as seen in Fig.
\ref{fig:Scattering}b.

We estimate the mean photon number per second integrated over the solid angle for the only two experiments so far on light diffraction from truly ultracold bosons where the measurement object was light
	\begin{equation}
		n_{\Phi}= \left(\frac{\Omega_0}{\Delta_a}\right)^2 \frac{\Gamma K}{8} (\langle\hat{n}^2\rangle-\langle\hat{n}\rangle^2),
	\end{equation}
where $\Omega_0=d_A E_0/\hbar$ and $\Gamma$ is the atomic relaxation rate.  The background signal should reach $n_\Phi \approx 10^6$ s$^{-1}$ in Ref. \cite{Weitenberg2011} (150 atoms in 2D), and $n_\Phi \approx 10^{11}$ s$^{-1}$ in Ref. \cite{KetterlePRL2011} ($10^5$ atoms in 3D).

%%%%%%%%%%%%%%%%%%%%%%%%%%%%%%%%%%%%%%%%%%%%%%%%%%%%%%%%%%%%%%%%%%%%%%%%%%%%%
%%%%%%%%%%%%%%%%%%%%%%%%%%%%%%%%%%%%%%%%%%%%%%%%%%%%%%%%%%%%%%%%%%%%%%%%%%%%%
%%
%%\section{QND measurement}
%%
%%%%%%%%%%%%%%%%%%%%%%%%%%%%%%%%%%%%%%%%%%%%%%%%%%%%%%%%%%%%%%%%%%%%%%%%%%%%%
%%%%%%%%%%%%%%%%%%%%%%%%%%%%%%%%%%%%%%%%%%%%%%%%%%%%%%%%%%%%%%%%%%%%%%%%%%%%%

\subsection{Matter-field interference measurements}

\begin{figure*}[htbp!]
	\begin{center}
		\includegraphics[width=\linewidth]{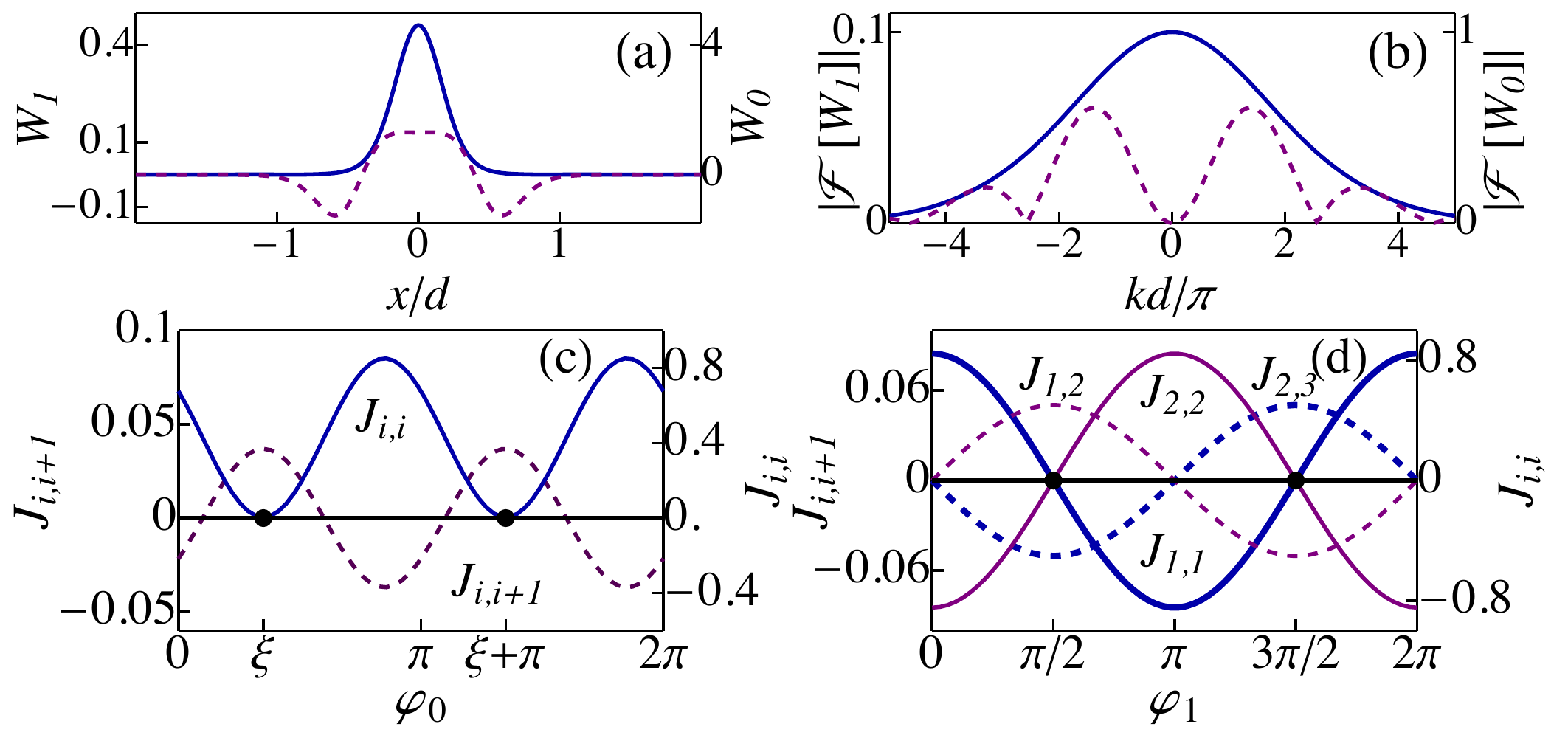}
	\end{center}
	\captionsetup{justification=centerlast,font=small}
	\caption{(Color online) The Wannier function products: (a) $W_0(x)$ (solid line, right axis), $W_1(x)$ (dashed line,  left axis) and their (b) Fourier transforms $\mathcal{F}[W_{0,1}]$. The Density $J_{i,i}$ and matter-interference $J_{i,i+1}$ coefficients  (\ref{eqDB}) in diffraction maximum (c) and minimum (d) as are shown as functions  of standing wave shifts $\varphi$ or, if one were to measure the quadrature variance $(\Delta X^F_\beta)^2$, the local oscillator phase  $\beta$. The black points indicate the positions, where light measures matter interference $\hat{B} \ne 0$, and the density-term is suppressed, $\hat{D} = 0$. The trapping potential depth is approximately 5 recoil energies.}
	\label{FJ}
\end{figure*}

We now focus on enhancing the interference term $\hat{B}$ in the operator $\hat{F}$. For clarity we will consider a 1D lattice, but the results can be applied and generalised to higher dimensions. Central to engineering the $\hat{F}$ operator are the coefficients $J_{i,j}$ given by Eq. (\ref{Jcoeff}). The operators $\hat{B}$ and $\hat{D}$ depend on the values of $J_{i,i+1}$ and $J_{i,i}$ respectively. These coefficients are determined by the convolution of the light mode product, $u_1^*({\bf r})u_0({\bf r})$ with the relevant Wannier function overlap shown in Fig. \ref{FJ}a. For the $\hat{B}$ operator we calculate the convolution with the nearest neighbour overlap, $W_1({\bf r}) \equiv w({\bf r} - {\bf d}/2) w({\bf r}+{\bf d}/2)$, and for the $\hat{D}$ operator we calculate the convolution with the square of the Wannier function at a single site, $W_0({\bf r}) \equiv w^2({\bf r})$. Therefore, in order to enhance the $\hat{B}$ term we need to maximise the overlap between the light modes and the nearest neighbour Wannier overlap, $W_1({\bf r})$. This can be achieved by concentrating the light between the sites rather than at atom positions. Ideally, one could measure between two sites similarly to single-site addressing \cite{GreinerNature2009,BlochNature2011}, which would measure a single term $\langle b^\dag_i b_{i+1}+b_i b^\dag_{i+1}\rangle$, e.g., by superposing a deeper OL shifted by $d/2$ with respect to the original one, catching and measuring the atoms in the new lattice sites. A single-shot success rate of atom detection will be small. As single-site addressing is challenging, we proceed with the global scattering.

In order to calculate the $J_{i,j}$ coefficients we perform numerical calculations using realistic Wannier functions \cite{Walters2013}. However, it is possible to gain some analytic insight into the behaviour of these values by looking at the Fourier transforms of the Wannier function overlaps, $\mathcal{F}[W_{0,1}]({\bf k})$ , shown in Fig \ref{FJ}b. This is because the light mode product, $u_1^*({\bf r})u_0({\bf r})$, can be in general decomposed into a sum of oscillating exponentials of the form $e^{i {\bf k} \cdot {\bf r}}$ making the integral in Eq. (\ref{Jcoeff}) a sum of Fourier transforms of $W_{0,1}({\bf r})$. We consider both the detected and probe beam to be standing waves which gives the following expressions for the $\hat{D}$ and $\hat{B}$ operators
\begin{eqnarray}
\label{FTs}
	\hat{D} = \frac{1}{2}[\mathcal{F}[W_0](k_-)\sum_m\hat{n}_m\cos(k_- x_m +\varphi_-)  \nonumber\\
	+\mathcal{F}[W_0](k_+)\sum_m\hat{n}_m\cos(k_+ x_m +\varphi_+)], \nonumber\\
	\hat{B} = \frac{1}{2}[\mathcal{F}[W_1](k_-)\sum_m\hat{B}_m\cos(k_- x_m +\frac{k_-d}{2}+\varphi_-)  \nonumber\\
	+\mathcal{F}[W_1](k_+)\sum_m\hat{B}_m\cos(k_+ x_m +\frac{k_+d}{2}+\varphi_+)],
\end{eqnarray}
where $k_\pm = k_{0x} \pm k_{1x}$, $k_{(0,1)x} = k_{0,1} \sin(\theta_{0,1}$), $\hat{B}_m=b^\dag_mb_{m+1}+b_mb^\dag_{m+1}$, and $\varphi_\pm=\varphi_0 \pm \varphi_1$. The key result is that the $\hat{B}$ operator is phase shifted by $k_\pm d/2$ with respect to the $\hat{D}$ operator since it depends on the amplitude of light in between the lattice sites and not at the positions of the atoms, allowing to decouple them at specific angles.

Firstly, we will use this result to show how one can probe $\langle \hat{B} \rangle$ which in MF gives information about the matter-field amplitude, $\Phi = \langle b \rangle$. The simplest case is to find a diffraction maximum where $J_{i,i+1} = J_B$. This can be achieved by crossing the light modes such that $\theta_0 = -\theta_1$ and $k_{0x} = k_{1x} = \pi/d$ and choosing the light mode phases such that $\varphi_+ = 0$. Fig. \ref{FJ}c shows the value of the $J_{i,j}$ coefficients under these circumstances. In order to make the $\hat{B}$ contribution to light scattering dominant we need to set $\hat{D} = 0$ which from Eq. (\ref{FTs}) we see is possible if $\varphi_0 = -\varphi_1 = \arccos[-\mathcal{F}[W_0](2\pi/d)/\mathcal{F}[W_0](0)]/2$. This arrangement of light modes maximizes the interference signal, $\hat{B}$, by suppressing the density signal, $\hat{D}$, via interference compensating for the spreading of the Wannier functions. Hence, by measuring the light quadrature we probe the kinetic energy and, in MF, the matter-field amplitude (order parameter) $\Phi$: $\langle \hat{X}^F_{\beta=0} \rangle = | \Phi |^2 \mathcal{F}[W_1](2\pi/d) (K-1)$.

\begin{figure}[htbp!]
	\begin{center}
  		\includegraphics[width=\linewidth]{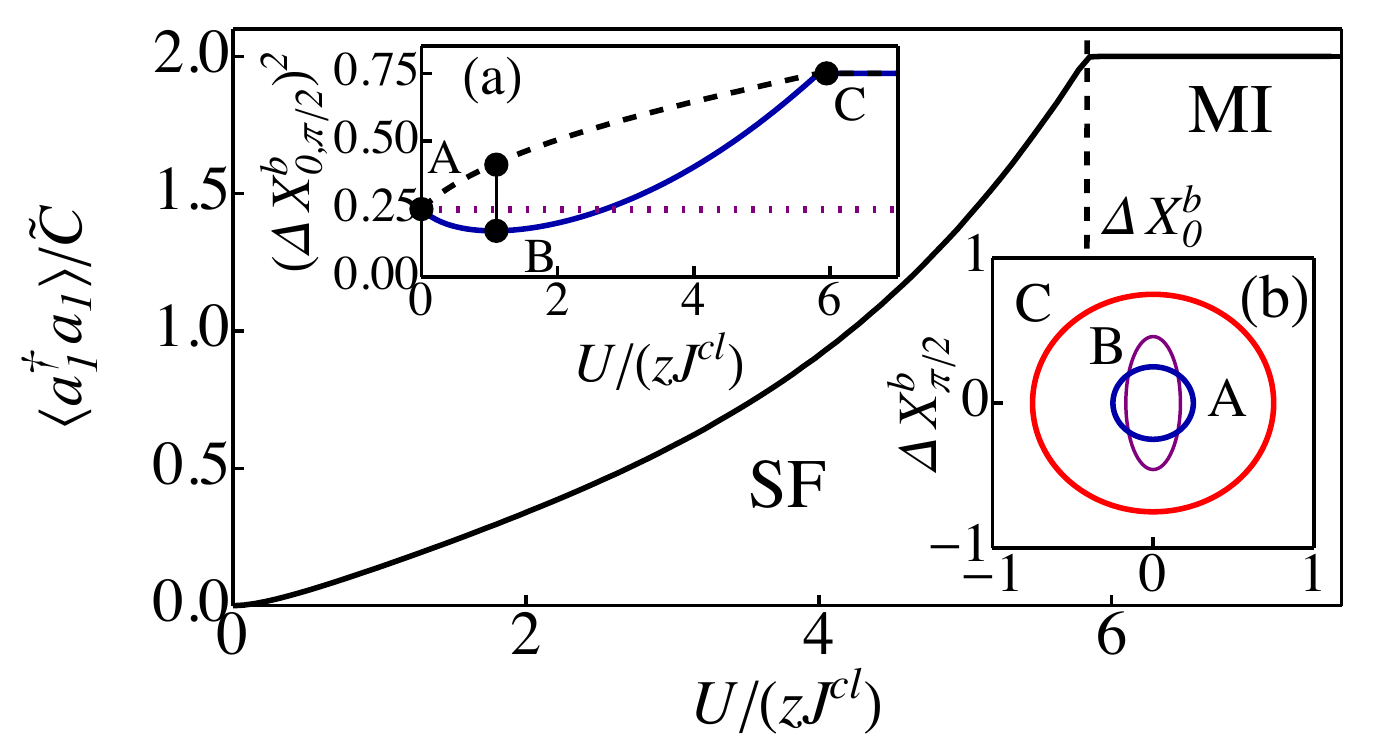}
 	\end{center}
	\captionsetup{justification=centerlast,font=small}
	\caption{(Color online) Photon number scattered in a diffraction minimum, given by Eq. (\ref{intensity}), where $\tilde{C} = 2 |C|^2 (K-1) \mathcal{F}^2 [W_1](\pi/d)$. 	More light is scattered from a MI than a SF due to the large uncertainty in phase in the insulator. (a) The variances of quadratures $\Delta X^b_0$ (solid) and $\Delta X^b_{\pi/2}$ (dashed) of the matter field across the phase transition. Level 1/4 is the minimal (Heisenberg) uncertainty. There are three important points along the phase transition: the coherent state (SF) at A, the amplitude-squeezed state at B, and the Fock state (MI) at C. (b) The uncertainties plotted in phase space.}
	\label{Quads}
\end{figure}

Secondly, we show that it is also possible to access the fluctuations of matter-field quadratures $\hat{X}^b_\alpha = (b e^{-i\alpha} + \bt e^{i\alpha})/2$, which in MF can be probed by measuring the variance of $\hat{B}$. Across the phase transition, the matter field changes its state from Fock (in MI) to coherent (deep SF) through an amplitude-squeezed state as shown in Fig. \ref{Quads}(a,b). We consider an arrangement where the beams are arranged such that $k_{0x} = 0$ and $k_{1x} = \pi/d$ which gives the following expressions for the density and interference terms
\begin{eqnarray}
\label{DMin}
	\hat{D} = \mathcal{F}[W_0](\pi/d) \sum_m (-1)^m \hat{n}_m \cos(\varphi_0) \cos(\varphi_1) \nonumber \\
	\hat{B} = -\mathcal{F}[W_1](\pi/d) \sum_m (-1)^m \hat{B}_m \cos(\varphi_0) \sin(\varphi_1).
\end{eqnarray}
The corresponding $J_{i,j}$ coefficients are shown in Fig. \ref{FJ}d for $\varphi_0=0$. It is clear that for $\varphi_1 = \pm \pi/2$, $\hat{D} = 0$, which is intuitive as this places the lattice sites at the nodes of the mode $u_1({\bf r})$. This is a diffraction minimum as the light amplitude is also zero, $\langle \hat{B} \rangle = 0$, because contributions from alternating inter-site regions interfere destructively. However, the intensity $\langle \at_1 a \rangle = |C|^2 \langle \hat{B}^2 \rangle$ is proportional to the variance of $\hat{B}$ and is non-zero. Assuming $\Phi$ is real in MF:
\begin{multline}
\label{intensity}
	\langle a_1^\dag a_1\rangle = 2 |C|^2(K-1)\mathcal{F}^2[W_1](\frac{\pi}{d}) \\
	\times [ ( \langle b^2 \rangle - \Phi^2 )^2 + ( n - \Phi^2 ) ( 1 +n - \Phi^2 ) ]
\end{multline}
and it is shown as a function of $U/(zJ^\text{cl})$ in Fig. \ref{Quads}. Thus, since measurement in the diffraction maximum yields $\Phi^2$ we can deduce $\langle b^2 \rangle - \Phi^2$ from the intensity. This quantity is of great interest as it gives us access to the quadrature variances of the matter-field 
\begin{equation}
	(\Delta X^b_{0,\pi/2})^2 = 1/4 + [(n - \Phi^2) \pm (\langle b^2 \rangle - \Phi^2)]/2,
\end{equation}
where $n=\langle\hat{n}\rangle$ is the mean on-site atomic density.

Alternatively, one can use the arrangement for a diffraction minimum described above, but use travelling instead of standing waves for the probe and detected beams and measure the light quadrature variance. In this case $\hat{X}^F_\beta = \hat{D} \cos(\beta) + \hat{B} \sin(\beta)$ and by varying the local oscillator phase, one can choose which conjugate operator to measure. For $\beta=\pi/2$, $(\Delta X^F_{\pi/2})^2$ looks identical to Eq. (\ref{intensity}).

Probing $\hat{B}^2$ gives us access to kinetic energy fluctuations with 4-point correlations ($\bt_i b_j$ combined in pairs). Measuring the photon number variance, which is standard in quantum optics, will lead up to 8-point correlations similar to 4-point density correlations \cite{Mekhov2007a}. These are of significant interest, because it has been shown that there are quantum entangled states that manifest themselves only in high-order correlations \cite{Kaszlikowski2008}.

Surprisingly, inter-site terms scatter more light from a MI than a SF Eq. (\ref{intensity}), as shown in Fig. (\ref{Quads}), although the mean inter-site density $\langle \hat{n}(\bf{r})\rangle $ is tiny in a MI. This reflects a fundamental effect of the boson interference in Fock states. It indeed happens between two sites, but as the phase is uncertain, it results in the large variance of $\hat{n}(\bf{r})$ captured by light as shown in Eq. (\ref{intensity}). The interference between two macroscopic BECs has been observed and studied theoretically \cite{Horak1999}. When two BECs in Fock states interfere a phase difference is established between them and an interference pattern is observed which disappears when the results are averaged over a large number of experimental realizations. This reflects the large shot-to-shot phase fluctuations corresponding to a large inter-site variance of $\hat{n}(\mathbf{r})$. By contrast, our method enables the observation of such phase uncertainty in a Fock state directly between lattice sites on the microscopic scale in-situ.

%%%%%%%%%%%%%%%%%%%%%%%%%%%%%%%%%%%%%%%%%%%%%%%%%%%%%%%%%%%%%%%%%%%%%%%%%%%%%
%%%%%%%%%%%%%%%%%%%%%%%%%%%%%%%%%%%%%%%%%%%%%%%%%%%%%%%%%%%%%%%%%%%%%%%%%%%%%
%%
%%\subsection{Numerical calculations}
%%
%%%%%%%%%%%%%%%%%%%%%%%%%%%%%%%%%%%%%%%%%%%%%%%%%%%%%%%%%%%%%%%%%%%%%%%%%%%%%
%%%%%%%%%%%%%%%%%%%%%%%%%%%%%%%%%%%%%%%%%%%%%%%%%%%%%%%%%%%%%%%%%%%%%%%%%%%%%

\subsection{Mapping the quantum phase diagram}

\begin{figure*}[htbp!]
	\centering
		\includegraphics[width=\linewidth]{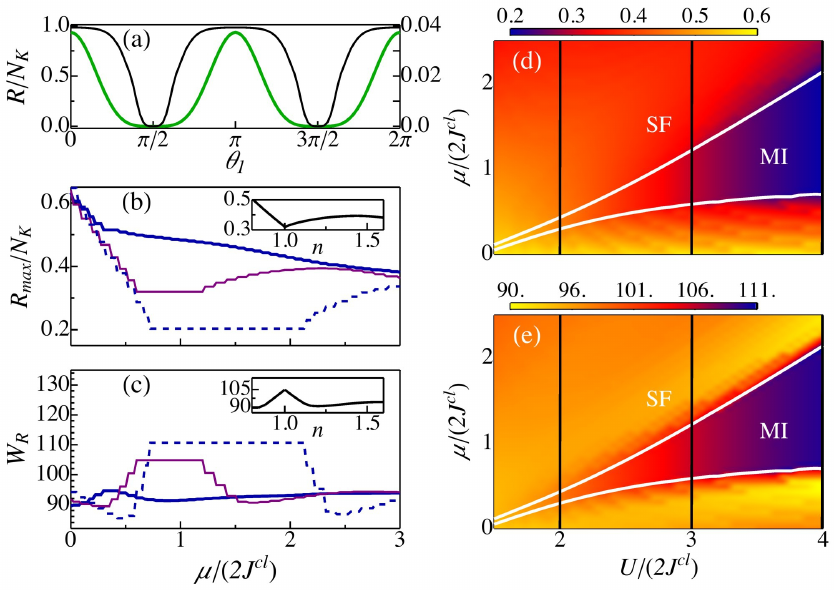}
	\captionsetup{justification=centerlast,font=small}
	\caption{(Color online) (a) The angular dependence of scattered light $R$ for SF (thin black, left scale, $U/2J^\text{cl} = 0$) and MI (thick green, right scale, $U/2J^\text{cl} =10$). The two phases differ in both their value of $R_\text{max}$ as well as $W_R$ showing that density correlations in the two phases differ in magnitude as well as extent. Light scattering maximum $R_\text{max}$ is shown in (b, d) and the width $W_R$ in (c, e).  It is very clear that varying chemical potential $\mu$ or density $\langle n\rangle$ sharply identifies the SF-MI transition in both quantities. (b) and (c) are cross-sections of the phase diagrams (d) and (e) at $U/2J^\text{cl}=2$ (thick blue), 3 (thin purple), and 4 (dashed blue). Insets show density dependencies for the $U/(2 J^\text{cl}) = 3$ line. $K=M=N=25$.}
	\label{fig:SFMI}
\end{figure*}

We have shown how in MF, we can track the order parameter, $\Phi$, by probing the matter-field interference using the coupling of light to the $\hat{B}$ operator. In this case, it is very easy to follow the SF-MI quantum phase transition since we have direct access to the order parameter which goes to zero in the insulating phase. In fact, if we're only interested in the critical point, we only need access to any quantity that yields information about density fluctuations which also go to zero in the MI phase and this can be obtained by measuring $\langle \hat{D}^\dagger \hat{D} \rangle$. However, there are many situations where the MF approximation is not a valid description of the physics. A prominent example is the BHM in 1D \cite{Cazalilla2011,Ejima2011, Kuhner2000, Pino2012, Pino2013}. Observing the transition in 1D by light at fixed density was considered to be difficult \cite{DeChiaraPRA2014} or even impossible \cite{BurnettPRA2003}. By contrast, here we propose to vary the density or chemical potential, which sharply identifies the transition. We perform these calculations numerically by calculating the ground state using DMRG methods \cite{tnt} from which we can compute all the necessary atomic observables. Experiments typically use an additional harmonic confining potential on top of the OL to keep the atoms in place which  means that the chemical potential will vary in space. However, with careful consideration of the full ($\mu/2J^\text{cl}$, $U/2J^\text{cl}$) phase diagrams in Fig. \ref{fig:SFMI}(d,e) our analysis can still be applied to the system \cite{Batrouni2002}.

The 1D phase transition is best understood in terms of two-point correlations as a function of their separation \cite{Giamarchi}. In the MI phase, the two-point correlations $\langle b_i^\dagger b_j \rangle$ and $\langle \delta \hat{n}_i \delta \hat{n}_j \rangle$ ($\delta \hat{n}_i =\hat{n}_i-\langle \hat{n}_i\rangle$) decay exponentially with $|i-j|$. On the other hand the SF will exhibit long-range order which in dimensions higher than one, manifests itself with an infinite correlation length. However, in 1D only pseudo long-range order happens and both the matter-field and density fluctuation correlations decay algebraically \cite{Giamarchi}.

The method we propose gives us direct access to the structure factor, which is a function of the two-point correlation $\langle \delta \hat{n}_i \delta \hat{n}_j \rangle$, by measuring the light intensity. For two travelling waves maximally coupled to the density (atoms are at light intensity maxima so $\hat{F} = \hat{D}$), the quantum addition is given by
\begin{equation}
	R =\sum_{i, j} \exp[i (\mathbf{k}_1 - \mathbf{k}_0) (\mathbf{r}_i - \mathbf{r}_j)] \langle \delta \hat{n}_i \delta \hat{n}_j \rangle,
\end{equation}

The angular dependence of $R$ for a MI and a SF is shown in Fig. \ref{fig:SFMI}a, and there are two variables distinguishing the states. Firstly, maximal $R$, $R_\text{max} \propto \sum_i \langle \delta \hat{n}_i^2 \rangle$, probes the fluctuations and compressibility $\kappa'$ ($\langle \delta \hat{n}^2_i \rangle \propto \kappa' \langle \hat{n}_i \rangle$).  The MI is incompressible and thus will have very small on-site fluctuations and it will scatter little light leading to a small $R_\text{max}$. The deeper the system is in the MI phase (i.e. that larger the $U/2J^\text{cl}$ ratio is), the smaller these values will be until ultimately it will scatter no light at all in the $U \rightarrow \infty$ limit. In Fig. \ref{fig:SFMI}a this can be seen in the value of the peak in $R$. The value $R_\text{max}$ in the SF phase ($U/2J^\text{cl} = 0$) is larger than its value in the MI phase ($U/2J^\text{cl} = 10$) by a factor of $\sim$25. Figs. \ref{fig:SFMI}(b,d) show how the value of $R_\text{max}$ changes across the phase transition. We see that the transition shows up very sharply as $\mu$ is varied.

Secondly, being a Fourier transform, the width $W_R$ of the dip in $R$ is a direct measure of the correlation length $l$, $W_R \propto 1/l$. The MI being an insulating phase is characterised by exponentially decaying correlations and as such it will have a very large $W_R$. However, the SF in 1D exhibits pseudo long-range order which manifests itself in algebraically decaying two-point correlations \cite{Giamarchi} which significantly reduces the dip in the $R$. This can be seen in Fig. \ref{fig:SFMI}a and we can also see that this identifies the phase transition very sharply as $\mu$ is varied in Figs. \ref{fig:SFMI}(c,e). One possible concern with experimentally measuring $W_R$ is that it might be obstructed by the classical diffraction maxima which appear at angles corresponding to the minima in $R$. However, the width of such a peak is much smaller as its width is proportional to $1/M$.

It is also possible to analyse the phase transition quantitatively using our method. Unlike in higher dimensions where an order parameter can be easily defined within the MF approximation there is no such quantity in 1D. However, a valid description of the relevant 1D low energy physics is provided by Luttinger liquid theory \cite{Giamarchi}. In this model correlations in the SF phase as well as the SF density itself are characterised by the Tomonaga-Luttinger parameter, $K_b$. This parameter also identifies the phase transition in the thermodynamic limit at $K_b = 1/2$. This quantity can be extracted from various correlation functions and in our case it can be extracted directly from $R$ \cite{Ejima2011}. By extracting this parameter from $R$ for various lattice lengths from numerical DMRG calculations it was even possible to give a theoretical estimate of the critical point for commensurate filling, $N = M$, in the thermodynamic limit to occur at $U/2J^\text{cl} \approx 1.64$ \cite{Ejima2011}. Our proposal provides a method to directly measure $R$ in a lab which can then be used to experimentally determine the location of the critical point in 1D.

So far both variables we considered, $R_\text{max}$ and $W_R$, provide similar information. Next, we present a case where it is very different. BG is a localized insulating phase with exponentially decaying correlations but large compressibility and on-site fluctuations in a disordered OL. Therefore, measuring both $R_\text{max}$ and $W_R$ will distinguish all the phases. In a BG we have finite compressibility, but exponentially decaying correlations. This gives a large $R_\text{max}$ and a large $W_R$. A MI will also have exponentially decaying correlations since it is an insulator, but it will be incompressible. Thus, it will scatter light with a small $R_\text{max}$ and large $W_R$. Finally, a SF will have long range correlations and large compressibility which results in a large $R_\text{max}$ and a small $W_R$.

\begin{figure}[htbp!]
	\centering
		\includegraphics[width=\linewidth]{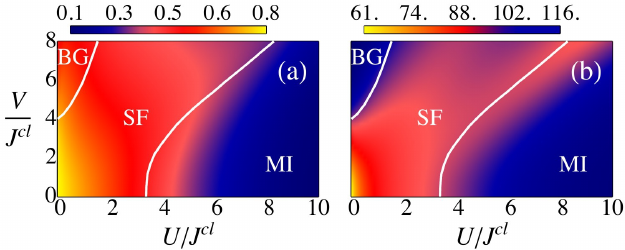}
	\captionsetup{justification=centerlast,font=small}
	\caption{(Color online) The MI-SF-BG phase diagrams for light scattering maximum $R_\text{max}/N_K$ (a) and width $W_R$ (b). Measurement of both quantities distinguish all three phases. Transition lines are shifted due to finite size effects \cite{Roux2008}, but it is possible to apply well known numerical methods to extract these transition lines from such experimental data extracted from $R$ \cite{Ejima2011}. $K=M=N=35$.}
	\label{fig:BG}
\end{figure}

We confirm this in Fig. \ref{fig:BG} for simulations with the ratio of superlattice- to trapping lattice-period $r\approx 0.77$ for various disorder strengths $V$ \cite{Roux2008}. Here, we only consider calculations for a fixed density, because the usual interpretation of the phase diagram  in the ($\mu/2J^\text{cl}$, $U/2J^\text{cl}$) plane for a fixed ratio $V/U$ becomes complicated due to the presence of multiple compressible and incompressible phases between successive MI lobes \cite{Roux2008}. This way, we have limited our parameter space to the three phases we are interested in: SF, MI, and BG. From Fig. \ref{fig:BG} we see that all three phases can indeed be distinguished. In the 1D BHM there is no sharp MI-SF phase transition in 1D at a fixed density \cite{Cazalilla2011,Ejima2011, Kuhner2000, Pino2012, Pino2013} just like in Figs. \ref{fig:SFMI}(d,e) if we follow the transition through the tip of the lobe which corresponds to a line of unit density. However, despite the lack of an easily distinguishable critical point it is possible to quantitatively extract the location of the transition lines by extracting the Tomonaga-Luttinger parameter from the scattered light, $R$, in the same way it was done for an unperturbed BHM \cite{Ejima2011}.

Only recently \cite{Derrico2014} BG was studied by combined measurements of coherence, transport, and excitation spectra, all of which are destructive techniques. Our method is simpler as it only requires measurement of the quantity $R$ and additionally, it is nondestructive.

%%%%%%%%%%%%%%%%%%%%%%%%%%%%%%%%%%%%%%%%%%%%%%%%%%%%%%%%%%%%%%%%%%%%%%%%%%%%%
%%%%%%%%%%%%%%%%%%%%%%%%%%%%%%%%%%%%%%%%%%%%%%%%%%%%%%%%%%%%%%%%%%%%%%%%%%%%%
%%
%%\subsection{Conclusions}
%%
%%%%%%%%%%%%%%%%%%%%%%%%%%%%%%%%%%%%%%%%%%%%%%%%%%%%%%%%%%%%%%%%%%%%%%%%%%%%%
%%%%%%%%%%%%%%%%%%%%%%%%%%%%%%%%%%%%%%%%%%%%%%%%%%%%%%%%%%%%%%%%%%%%%%%%%%%%%

\section{Conclusions}

In summary, we proposed a nondestructive method to probe quantum gases in an OL. Firstly, we showed that the density-term in scattering has an angular distribution richer than classical diffraction, derived generalized Bragg conditions, and estimated parameters for the only two relevant experiments to date \cite{Weitenberg2011,KetterlePRL2011}. Secondly, we proposed how to measure the matter-field interference by concentrating light between the sites. This corresponds to interference at the shortest possible distance in an OL. By
contrast, standard destructive time-of-flight measurements deal with far-field interference and a relatively near-field one was used in
Ref. \cite{KetterlePRL2011}. This defines most processes in OLs. E.g. matter-field phase changes may happen not only due to external
gradients, but also due to intriguing effects such quantum jumps leading to phase flips at neighbouring sites and sudden cancellation of
tunneling \cite{VukicsNJP2007}, which should be accessible by our method. In MF, one can measure the matter-field amplitude (order
parameter), quadratures and squeezing. This can link atom optics to areas where quantum optics has already made progress, e.g., quantum
imaging \cite{GolubevPRA2010,KolobovRMP1999}, using an OL as an array of multimode nonclassical matter-field sources with a high degree of entanglement for QIP. Thirdly, we demonstrated how the method accesses effects beyond MF and distinguishes all the phases in the MI-SF-BG transition, which is currently a challenge \cite{Derrico2014}. Based on off-resonant scattering, and thus being insensitive to a detailed atomic level structure, the method can be extended to molecules \cite{MekhovLP13}, spins, and fermions \cite{RuostekoskiPRL2009}.

\begin{acknowledgements}
The authors are grateful to D. Jaksch, S. Al-Assam, S. Clark, T. Johnson, E. Owen and EPSRC (DTA and EP/I004394/1).
\end{acknowledgements}

\bibliography{Kozlowski}

\end{document}